\documentclass[aps,twocolumn,showpacs,superscriptaddress]{revtex4}
\usepackage{epsfig}
\usepackage{times}
\usepackage{amsmath}

\begin{document}

\title{From degree-correlated to payoff-correlated activity for an optimal resolution of social dilemmas}

\author{Alberto Aleta}
\email{albertoaleta@gmail.com}
\affiliation{Institute for Biocomputation and Physics of Complex Systems, University of Zaragoza, E-50018 Zaragoza, Spain}

\author{Sandro Meloni}
\affiliation{Institute for Biocomputation and Physics of Complex Systems, University of Zaragoza, E-50018 Zaragoza, Spain}
\affiliation{Department of Theoretical Physics, University of Zaragoza, E-50009, Zaragoza, Spain}

\author{Matja{\v z} Perc}
\affiliation{Faculty of Natural Sciences and Mathematics, University of Maribor, Koro{\v s}ka cesta 160, SI-2000 Maribor, Slovenia}
\affiliation{CAMTP -- Center for Applied Mathematics and Theoretical Physics, University of Maribor, Krekova 2, SI-2000 Maribor, Slovenia}

\author{Yamir Moreno}
\affiliation{Institute for Biocomputation and Physics of Complex Systems, University of Zaragoza, E-50018 Zaragoza, Spain}
\affiliation{Department of Theoretical Physics, University of Zaragoza, E-50009, Zaragoza, Spain}
\affiliation{Complex Networks and Systems Lagrange Lab, Institute for Scientific Interchange, Turin 10126, Italy}

\begin{abstract}
An active participation of players in evolutionary games depends on several factors, ranging from personal stakes to the properties of the interaction network. Diverse activity patterns thus have to be taken into account when studying the evolution of cooperation in social dilemmas. Here we study the weak prisoner's dilemma game, where the activity of each player is determined in a probabilistic manner either by its degree or by its payoff. While degree-correlated activity introduces cascading failures of cooperation that are particularly severe on scale-free networks with frequently inactive hubs, payoff-correlated activity provides a more nuanced activity profile, which ultimately hinders systemic breakdowns of cooperation. To determine optimal conditions for the evolution of cooperation, we introduce an exponential decay to payoff-correlated activity that determines how fast the activity of a player returns to its default state. We show that there exists an intermediate decay rate, at which the resolution of the social dilemma is optimal. This can be explained by the emerging activity patterns of players, where the inactivity of hubs is compensated effectively by the increased activity of average-degree players, who through their collective influence in the network sustain a higher level of cooperation. The sudden drops in the fraction of cooperators observed with degree-correlated activity therefore vanish, and so does the need for the lengthy spatiotemporal reorganization of compact cooperative clusters. The absence of such asymmetric dynamic instabilities thus leads to an optimal resolution of social dilemmas, especially when the conditions for the evolution of cooperation are strongly adverse.
\end{abstract}

\maketitle

\section{Introduction}
The application of statistical physics to problems in evolutionary game theory has proven rewarding and conducive to inspiring results with a broad range of applicability \cite{szabo_pr07, roca_plr09, perc_bs10, perc_jrsi13, pacheco_plrev14, wang_z_epjb15, wang2015universal, szabo_pr16}. The successful evolution of cooperation in the realm of social dilemmas --- when what is best for an individual is at odds with what is best for the society --- is a subject that benefited particularly from this development, with innovative research revealing key mechanisms that can either promote or hinder the evolution of socially preferable states \cite{zimmermann_pre04, zimmermann_pre05, santos_pnas06, fu_pre09, du_wb_epl09, lee_s_prl11, gomez-gardenes_epl11, tanimoto_pre12, santos_md_srep14, pavlogiannis_srep15, wu_zx_epl15, hindersin_pcbi15, liu_rr_epl15, chen_w_pa16}. More specifically, it has been shown that phase transitions towards cooperative states depend on the properties of the interaction network and the strength of ties, as well as on the number and type of competing strategies \cite{santos_prl05, gomez-gardenes_prl07, assaf_prl12, xu_b_srep15, javarone_csn15, perc2016phase}.

The study of evolutionary dynamics and social dilemmas using methods borrowed from the physical sciences has been to a large degree motivated by the seminal work of Nowak and May \cite{nowak_n92b}, who showed that spatial structure might promote the evolution of cooperation through a mechanism that is referred to as network reciprocity \cite{nowak_s06}. During the last decade, independent research from different groups demonstrated the importance of heterogeneity of agents for the successful evolution of cooperation, be it introduced in the form of heterogeneous interaction networks -- note, however, that for the case of human interactions, it is not clear whether network reciprocity plays a key role \cite{gracia-lazaro_srep12, gracia-lazaro_pnas12} -- noisy disturbances to payoffs, or other player-specific properties like the teaching activity or the propensity to acquire new links over time \cite{santos_prl05, perc_njp06a, tanimoto_pre07b, szolnoki_epl07, perc_pre08, santos_n08, szolnoki_epl08, szolnoki_epjb08, jiang_ll_pre09, shigaki_epl12, hauser_jtb14, tanimoto2015network, javarone_epjb16, amaral2016evolutionary, matsuzawa2016spatial, javarone2016role, javarone2016conformity}. Evidently, the consideration of various sources of heterogeneity is crucial to the subject, especially concerning the subject of human cooperation \cite{perc2016phase, rand_tcs13, rand_pnas14}, since our societies are not made up of uniform individuals. On the contrary, inequalities in wealth, reputation, the number of friends, and many other measurables characterizing real life, abound.

Motivated by the above considerations, we study the evolution of cooperation in social dilemmas, where players differ in their activity. More precisely, we no longer assume that players always participate in each round of the game \cite{perra2012activity}, but rather, that their participation is probabilistic, with the probability depending on either their degree in the network or their current payoff. We thus obtain a setting with degree-correlated or payoff-correlated activity patterns, depending on which criterion is used. Importantly, inactive players, i.e., players that do not participate in a particular instance of the game, get a null payoff and are unable to replicate. In this sense, our consideration of inactivity is different from the consideration of loners \cite{szabo_prl02}, who can replicate, and who refuse to participate by default in exchange for a small but fixed income. The treatment of inactivity as a dynamically changing state of players rather than a strategy, reveals the fascinating complexity evoked by different activity patterns of players. As we will show, while degree-correlated activity introduces cascading failures of cooperation that can only be compensated by a tedious spatiotemporal reorganization of compact cooperative clusters, payoff-correlated activity provides a more nuanced activity profile that hinders systemic breakdowns of cooperation and provides optimal conditions for cooperators to survive if the exponential decay to baseline activity is appropriately adjusted.

The continuation of this paper is organized as follows. In Section II we first present the details of the studied mathematical model, while in Section III we presented the main results of our research. We conclude in Section IV with a discussion of the broader implications of our conclusions.

\section{Mathematical model}

We use the simplified version of the prisoner's dilemma game, where the key aspects of the social dilemma are preserved but its strength is determined by a single parameter \cite{nowak_n92b}. In particular, mutual cooperation yields the reward $R=1$, mutual defection leads to punishment $P=0$, and the mixed choice gives the cooperator the sucker's payoff $S=0$ and the defector the temptation $T>1$. We note that the selection of this widely used and representative parametrization gives results that remain valid in a broad range of pairwise social dilemmas, including the snowdrift and the stag-hunt game.

At this point, we note that the consideration of different activity patterns does not affect the expected outcome of social dilemmas in well-mixed populations, so that we thus focus on the evolutionary outcomes in structured populations. As the interaction network, we use scale-free networks generated according to the uncorrelated configuration model \cite{catanzaro_pre05}, each with an average degree $k=4.5$ and size $N=10^{4}$. Cooperators ($C$) and defectors ($D$) are initially distributed uniformly at random, so that they have equal chances for success during the evolution.

Simulations are performed in agreement with a synchronous updating protocol, such that each player $i$ plays the game with all its $k_i$ neighbors and thereby collects the payoff $\Pi_i$. Once all the players collected their payoffs an evolutionary step takes place. For the evolution of strategies, we employ the replicator dynamics, such that if $\Pi_j > \Pi_i$, player $j$ will replicate its strategy $s_j$ to the site of player $i$ with the probability
\begin{equation}
P_{s_j \rightarrow s_i}= \frac{\Pi_j-\Pi_i}{\mbox{max}(k_j,k_i)b}.
\end{equation}
We perform simulations until the average fraction of cooperators in the population $\langle c \rangle$ reaches its long time pattern $-$note that it could be an oscillatory or a steady state.

To introduce different activity patterns, we assign to every player an activation probability $P_{a_i} \in [0,1]$ according to which player $i$ participates in a particular round of the game. Accordingly, with probability $1-P_{a_i}$ the player $i$ remains inactive in each round of the game. If a player $i$ is active, it will play the game with all its $k_i$ neighbors, independently from whether they are active or not, and it will obtain the payoff $\Pi_i$ as dictated by all the strategy pairs. If a player $i$ is inactive, it will not play any games with its neighbors, and accordingly, its payoff $\Pi_i$ will be null. Furthermore, only active players are able to pass their strategy in agreement with the replicator equation, while inactive players are never able to replicate because their payoff is always zero (and hence never larger than the payoff of the neighbor).

\begin{figure*}
\centerline{\epsfig{file=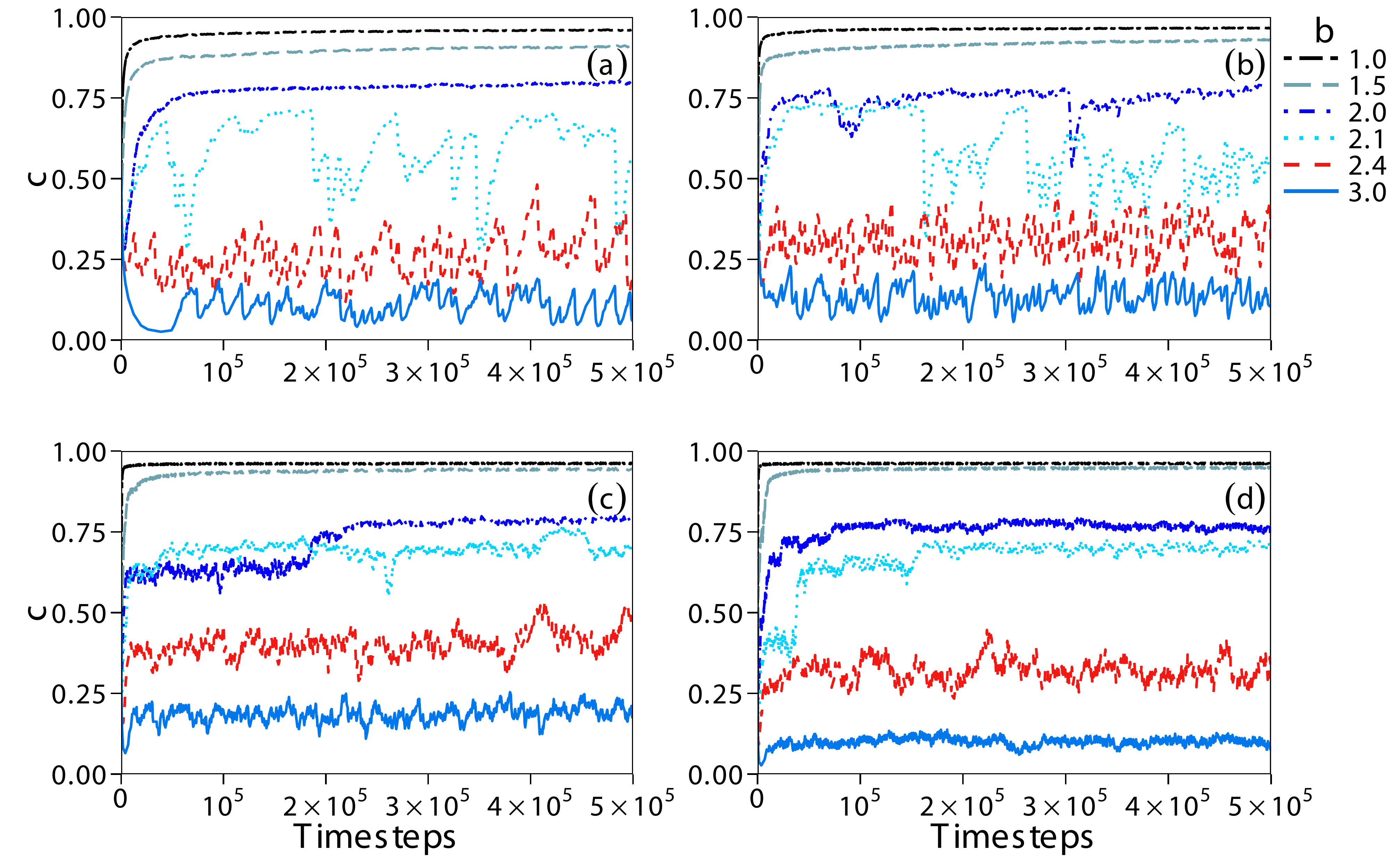,width=11.5cm}}
\caption{Time evolution of the fraction of cooperators, as obtained for different values of $b$ (see legend) and with different activity rules. Results obtained with degree-correlated activity are depicted in the top left panel (a), while results obtained with payoff-correlated activity with $\tau=1$, $4$ and $8$ are depicted in the top right and bottom two panels, (b), (c) and (d), respectively. It can be observed, particularly for $b=2.1$, that the sharp drops in cooperation vanish when going from degree-correlated to payoff-correlated activity with increasingly large values of $\tau$. At the same time, however, the plateau level of the fraction of cooperators drops as well too.}
\label{time}
\end{figure*}

Since the introduction of probabilistic participation of players in each instance of the game introduces a degree of heterogeneity to the evolutionary time scales and adds a layer of stochasticity to the spatiotemporal dynamics, it affects the correlation scales and can induce cascading effects that disrupt the formation of cooperative clusters. The distribution of activation probabilities in the population therefore plays a crucial role. Accordingly, we consider separately degree-correlated activity such that $P_{a_i} = k_i/k_{max}$ where $k_{max}$ is the largest degree in the network \cite{xia2015dynamic}, as well as payoff-correlated activity, such that
\begin{widetext}
\begin{equation}
P_{a_i}(t) = P_{a_i}(t-1) + \frac{1}{\alpha}f(P_{a_i},\pi_i)\frac{\pi_i(t-1) - \langle \pi_i \rangle_{(t-2,\ldots,t-k_i)}}{bk_i} - \Delta P_{a_i} e^{-\tau (1-|\Delta P_{a_i}|)}
\label{payc}
\end{equation}
\end{widetext}
\noindent where $\Delta P_{a_i} = P_{a_i}(t-1) - P_{a_i}(0)$, $P_{a_i}(0)=k_i/k_{\text{max}}$, and $\langle \pi_i \rangle_{(t-2,\ldots,t-k_i)}$ represents the weighted mean of the payoff obtained by node $i$ over the last $k_i$ time steps with weights given by $w_j=(k_i-j)/k_i$. Lastly, $f(P_{a_i},\pi_i)$ is equal to
\begin{widetext}
\begin{equation}
f(P_{a_i},\pi_i) = \left\{
\begin{array}{ll} 1-P_{a_i}(t-1) & \text{if } \pi_i(t-1)> \langle\pi_i\rangle_{(t-2,\ldots,t-k_i)} \\P_{a_i}(t-1) & \text{if } \pi_i(t-1) < \langle \pi_i \rangle_{(t-2,\ldots,t-k_i)}  \\
0 & \text{otherwise} \end{array} \right.
\label{fpa}
\end{equation}
\end{widetext}
\noindent The last term of Eq.~(\ref{payc}) depends only on the difference between the current activation probability and the initial activation probability at time $t=0$. Thus, the bigger the difference between the current and the initial activation probability, and the larger the value of $\tau$, the faster the activity of a player will return to its original state. The second term of of Eq.~(\ref{payc}) represents the variation that is due to the payoff obtained by each player. We first compute the average payoff obtained by each player over the last $k_i$ time steps so that a player has the longer memory the higher its degree. To increase the importance of the last rounds of the game we weight the average $w_j = (k_i-j)/k_i$ with
\begin{equation}
\langle \pi_i (t) \rangle_{(t-1,\ldots,t-k_i)} = \frac{\displaystyle\sum\limits_{j=0}^{k_i-1} w_j \pi_i (t-1-j)}{\displaystyle\sum\limits_{j=0}^{k_i-1} w_j}
\label{fin}
\end{equation}
\noindent and then compute the difference between the payoff obtained in the last round and this average, and finally normalize it with the factor $bk_i$. In the limit where $\langle \pi_i \rangle \rightarrow 0$ and $\pi_i(t-1) \rightarrow bk_i$ this fraction would take the value of $1$. The function $f(P_{a_i},\pi_i)$, given by Eq.~(\ref{fpa}), which multiplies the second term in Eq.~(\ref{payc}), represents the difference between the current activation probability and its upper limit if the payoff increases and between the current activation probability and its lower limit if the payoff decreases. Thus, in the limit where the previous fraction takes the value of $1$ this term will try to set the activation probability to $1$, whereas if it takes a value of $-1$ this term will try to set it to $0$. As the described change can be quite abrupt, we include a parameter $\alpha$ to modulate this behavior. In the following, we will consider $\alpha = \tau$ without loss of generality, thus leaving us with $\tau$ as the key parameter determining the interplay between the linear term and the exponential term in Eq.~(\ref{payc}) when considering payoff-driven activity of players. 

We note that this mechanism is similar to the Pavlov strategy (also known as win-stay-lose-shift \cite{nowak_nature93, chen_xj_pa08, liu_yk_pone12, amaral2016stochastic}), but adapted to the activation probability. Indeed, a player will tend to keep playing as long as it is winning, but will diminish his activity if it loses. This provides a game-exit option that was not included in previous models \cite{xia2015dynamic}.

\section{Results}

\begin{figure*}
\centerline{\epsfig{file=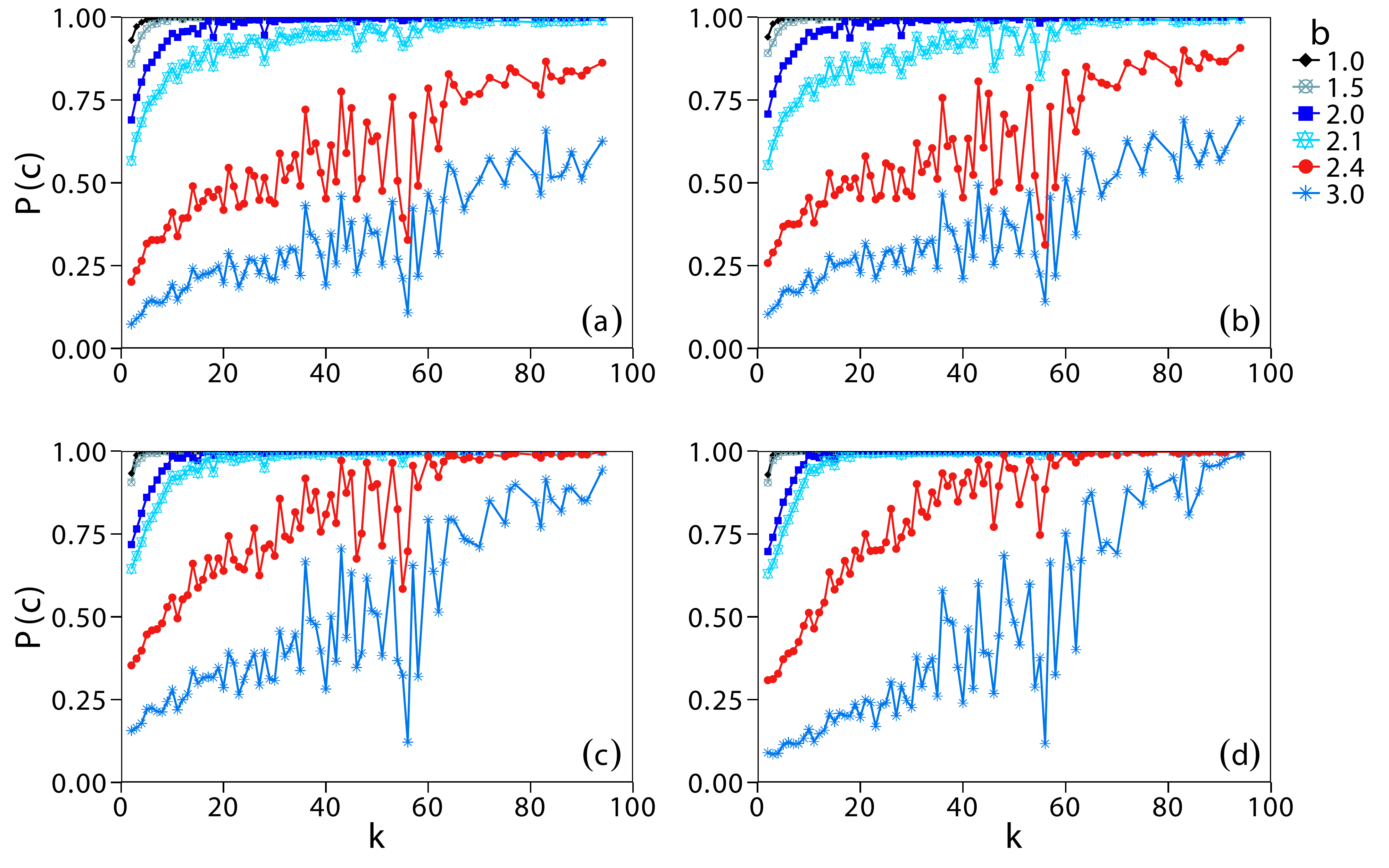,width=11.5cm}}
\caption{Cooperation probability in dependence on the degree of players, as obtained for different values of $b$ (see legend) and with different activity rules. Results obtained with degree-correlated activity are depicted in the top left panel (a), while results obtained with payoff-correlated activity with $\tau=1$, $4$ and $8$ are depicted in the top right and bottom two panels, (b), (c) and (d), respectively. The shift towards more cooperative hubs and more cooperative average-degree players is evident when going from degree-correlated to payoff-correlated activity, especially for $\tau=4$ and $b>2$.}
\label{prob}
\end{figure*}

Based on the consideration of limiting cases entering Eqs.~(\ref{payc}-\ref{fin}) and the shape of the resulting $P_{a_i}(t)$ function, the most interesting interval to explore in terms of the exponential decay that determines how fast the activity of a player returns to its default state is $1\lesssim \tau \lesssim 10$. We thus begin by showing representative time courses of the fraction of cooperators $c$, as obtained for different values of the temptation to defect $b$, and for different activity rules, in Fig.~\ref{time}. For degree-correlated activity, and since the game is staged on a scale-free network, the cascading failures of cooperation, which are due to frequently inactive hubs, are clearly observable, especially for $b=2.1$. On Erd\H{o}s-R\'enyi networks, on the other hand, the activity pattern does not induce sudden drops of the fraction of cooperators due to the lack of hubs, and thus cooperators rely predominantly on traditional network reciprocity \cite{xia2015dynamic}. Also, it can be observed that the level of cooperation on a scale-free network requires a comparatively long time to recover from the sudden drops, which is due to the fact that the spatiotemporal reorganization of compact cooperative clusters requires time. Interestingly, not all values of $b$ expose this feature equally well, which has to do with the fact that scale-free networks are very potent amplifiers of cooperation \cite{xia2015dynamic}. Accordingly, it is only when the cooperation level starts to drop significantly in the population do the cascading failures become most pronounced. As $b$ is increased further, the overall level of cooperation becomes so low that again the cascading failures are less visible. For the sake of the clarity of our arguments, we thus focus on those values of $b$ that convey the crucial point most clearly.

For payoff-correlated activity with $\tau=1$ the evolutionary outcome is much the same as for degree-correlated activity, with common sharp drops in the level of cooperation, and relatively long recovery periods being clearly visible for $b=2.0$ and $2.1$ (see upper right panel in Fig.~\ref{time}). However, for $\tau=4$ and $\tau=8$ the sudden drops in the level of cooperation progressively vanish, and overall, the time courses become smoother. A qualitative change in the evolutionary dynamics when going to degree-correlated to payoff-correlated activity of players is thus observable. By comparing results obtained with $\tau=4$ and $\tau=8$ more closely, one can also infer that, in addition to the drops vanishing, the plateau values of the fraction of cooperators $c$ also become lower for the larger $\tau$ value. Accordingly, it is not immediately clear whether the absence of drops offsets the drops in the plateau values, and thus, whether payoff-correlated activity has any evolutionary advantages to degree-correlated activity.

\begin{figure}
\centerline{\epsfig{file=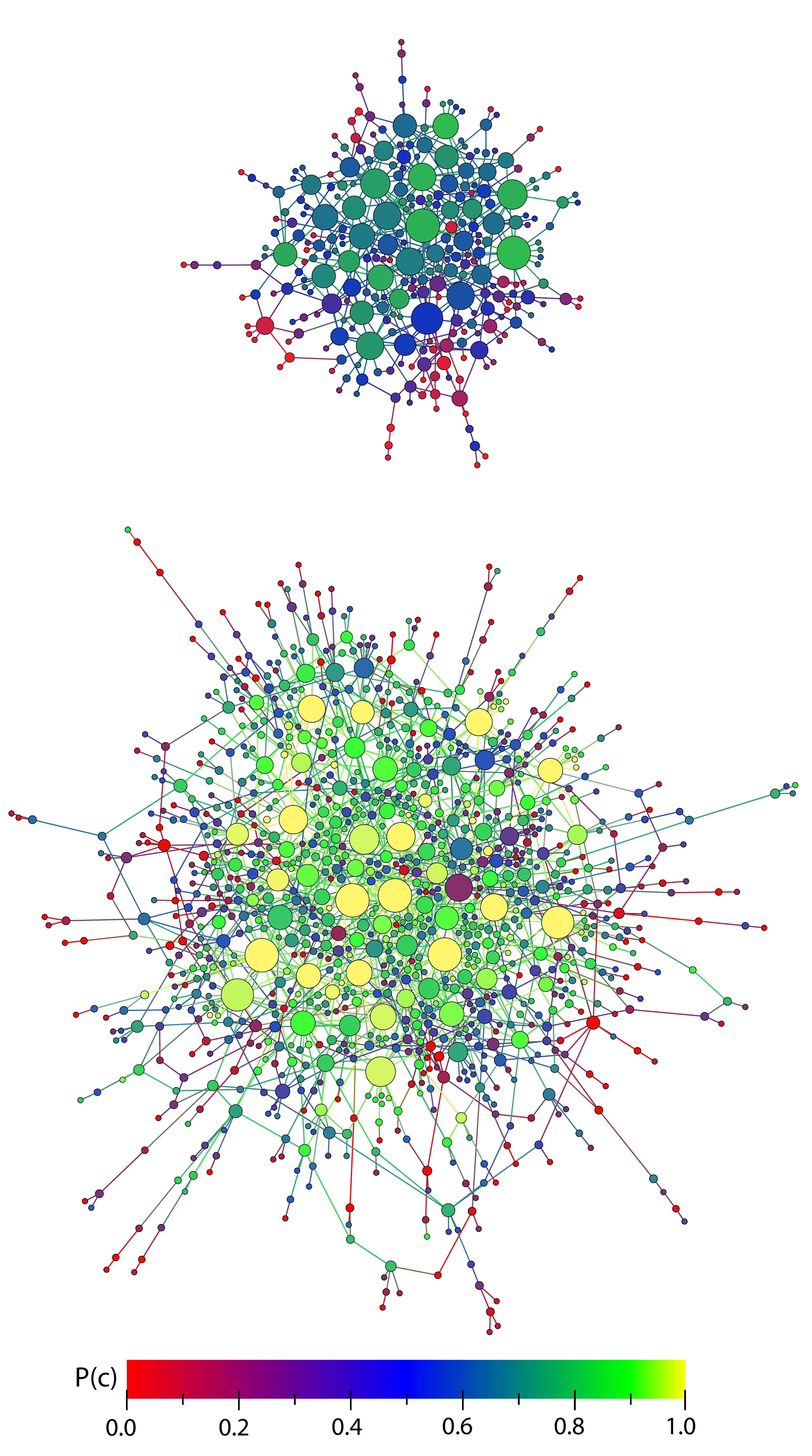,width=8cm}}
\caption{Snapshot of the giant component of the network, as obtained using degree-correlated (top) and payoff-correlated (bottom) activity at $b=2.4$ and $\tau=4$. The size of the players is proportional to their degree in the complete network, and their color is proportional to their cooperation probability as follows: red for defective nodes ($P_c\sim0$), blue for nodes with intermediate cooperation probability ($P_c\sim0.5$), green for highly cooperating nodes ($P_c\lesssim 0.9$), and yellow for almost pure cooperators ($P_c > 0.9$). Evidently, yellow color is more common among high-degree and average-degree nodes in the bottom network, where payoff-correlated activity is applied.}
\label{nets}
\end{figure}

To shed light on the details behind the results presented in Fig.~\ref{time}, we show in Fig.~\ref{prob} the cooperation probability in dependence on the degree of players forming the scale-free network. While for $b<2$ the probability profiles differ insignificantly across the panels, for $b=2.4$ the shift towards more cooperative hubs and more cooperative average-degree players is evident when going from degree-correlated to payoff-correlated activity with $\tau=1,4$ and $8$. However, while for $\tau=1$ and $\tau=4$ low degree nodes still maintain a relatively high cooperation probability too, for $\tau=8$ the same probability tappers off significantly faster. Accordingly, as $\tau$ increases hubs and average-degree nodes increase their chances of retaining a cooperative state, but low-degree nodes see their chances for cooperation diminished. A clear tradeoff thus appears to emerge between keeping the hubs and average-degree nodes cooperating, while at the same time not loosing low-degree nodes to defection. In this sense one can already envisage an optimal value of $\tau$ that should work best for the resolution of severe social dilemmas.

\begin{figure}
\centerline{\epsfig{file=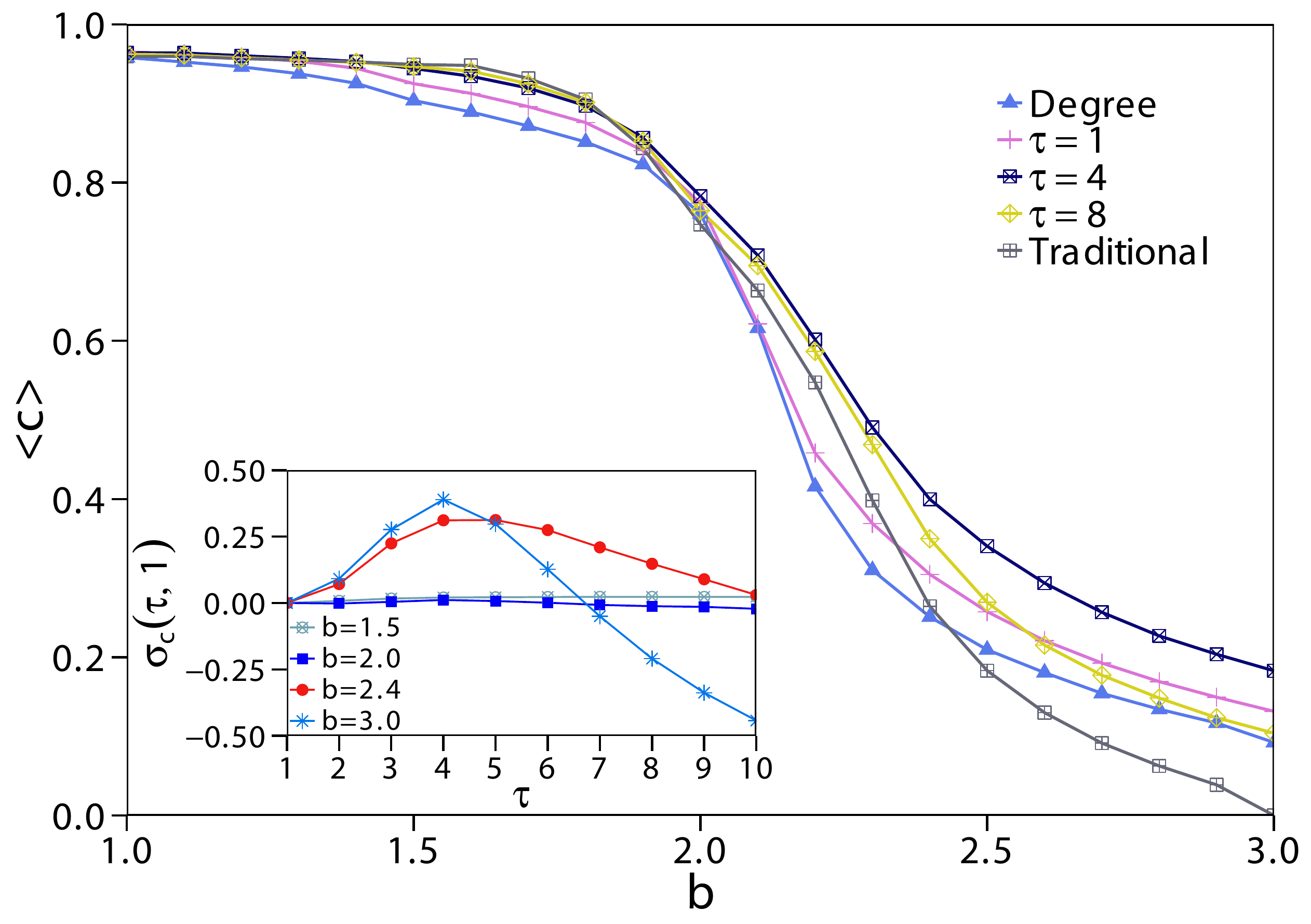,width=8.5cm}}
\caption{Average fraction of cooperators in dependence on the temptation to defect $b$, as obtained for the traditional version of the game with a uniform activity profile (TD), with degree-correlated activity (degree), and with payoff-correlated activity ($\tau=1,4,8$). The inset shows the  normalized difference of the average fraction of cooperators (see main text for details) in dependence on the exponential decay rate to baseline activity $\tau$. Both results in the main panel and the inset show that there exists an optimal value of $\tau$ at which, particularly for $b>2$ the level of cooperation in the population is highest.}
\label{optimal}
\end{figure}

Before making these argument quantitatively more precise, it is also instructive to examine representative snapshots of the network, as shown in Fig.~\ref{nets}, where players are color-coded in agreement with their cooperation probabilities. Also, the size of each player corresponds to its degree in the network (log-scale). The color code is such that red is used for defectors ($P_c\sim0$), blue for players with intermediate cooperation probability ($P_c\sim0.5$), green for very likely cooperators ($P_c\lesssim 0.9$), and yellow for virtually pure cooperators ($P_c > 0.9$). It can be observed at a glance that for payoff-correlated activity with $\tau=4$ (lower panel) the hubs and average-degree nodes are predominantly yellow, while for degree-correlated activity the hubs are green at best. There is also lots of red color, i.e., defectors, on the periphery of both depicted giant components, yet in neither cases this extends significantly to larger nodes, i.e., players with a larger degree within the network. The depicted two snapshots thus confirm that an appropriately set value of $\tau$ can effectively increase the probability for hubs and average-degree nodes to maintain a cooperative state, while at the same time not negatively affecting the cooperation probability for lower-degree nodes. As observed in Fig.~\ref{prob}, the latter does not necessarily hold if the value of $\tau$ increases, as then low-degree players in the network see their cooperation probability significantly lowered.

\begin{figure*}
\centerline{\epsfig{file=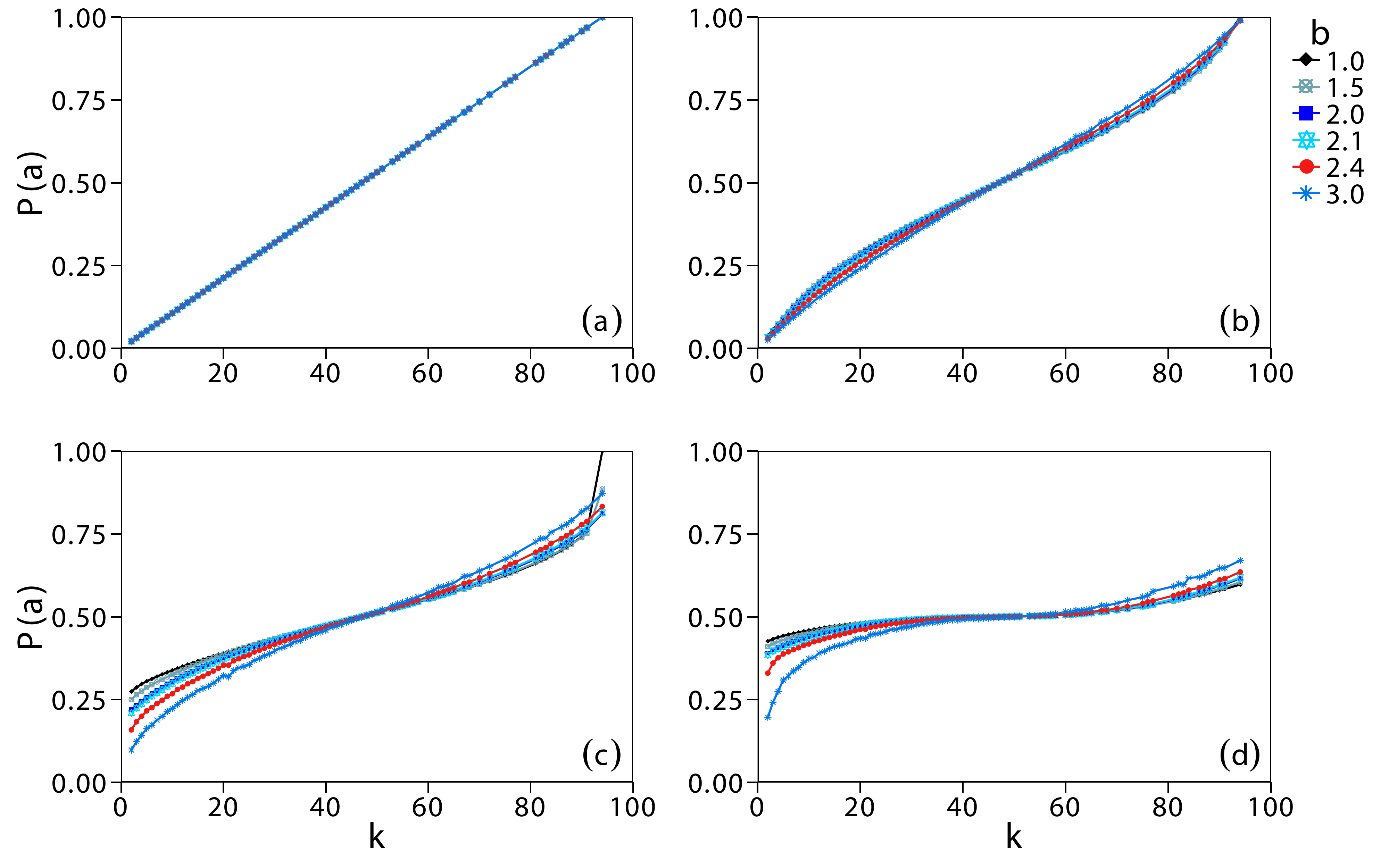,width=11.5cm}}
\caption{Activation probability in dependence on the degree of players, as obtained for different values of $b$ (see legend) and with different activity rules. Results obtained with degree-correlated activity are depicted in the top left panel (a), while results obtained with payoff-correlated activity with $\tau=1$, $4$ and $8$ are depicted in the top right and bottom two panels, (b), (c) and (d), respectively. The shift from a linear interpolation of activity for the degree-correlated case to an S-shaped distribution as $\tau$ increases is clearly visible, especially for $b>2$. Note that for the degree-correlated case $P_{a_i} = k_i/k_\text{max}$, and thus $P(a)$ is independent of $b$.}
\label{activate}
\end{figure*}

Results presented in Fig.~\ref{optimal} make our arguments made so far quantitative, showing in the main panel how the average fraction of cooperators in the network $\langle c\rangle$ varies in dependence on the temptation to defect $b$ for the traditional case with uniform activity (Traditional) and different activity patterns. It can be observed that, especially for $b>2$, where, as we have explained above the effects of different activity patterns are most clearly visible in Fig.~\ref{time}, the payoff-correlated activity with $\tau=4$ delivers consistently best results in comparison to all the other depicted cases. At the extreme end of the spectrum in terms of the severity of the social dilemma, for $b=3$, cooperators die out in the traditional case and maintain a $10 \%$ existence for degree-correlated activity, while for payoff-correlated activity with $\tau=4$ up to $20 \%$ of the population is cooperating. Moreover, the inset in Fig.~\ref{optimal} shows $\sigma_c (\tau,1) = \frac{\langle c(\tau)\rangle-\langle c(1)\rangle}{\langle c(1)\rangle}$ in dependence on $\tau$ for different values of $b$ to emphasize the existence of an optimal speed of the exponential decay to baseline activity of players under the payoff-driven rule. In particular, while the optimum for $b=1.5$ is absent, for $b=2.0$ is becomes faintly inferrable, and for $b=2.4$ and $3.0$ it is very clear, with the optimal value of $\tau$ being $4$ and $5$, respectively. This thus confirms the existence of the tradeoff between keeping the hubs and average-degree nodes cooperating, while at the same time not loosing low-degree nodes to defection, which combined for an optimal value of $\tau$ can lead to an optimal resolution of social dilemmas.

\begin{figure*}
\centerline{\epsfig{file=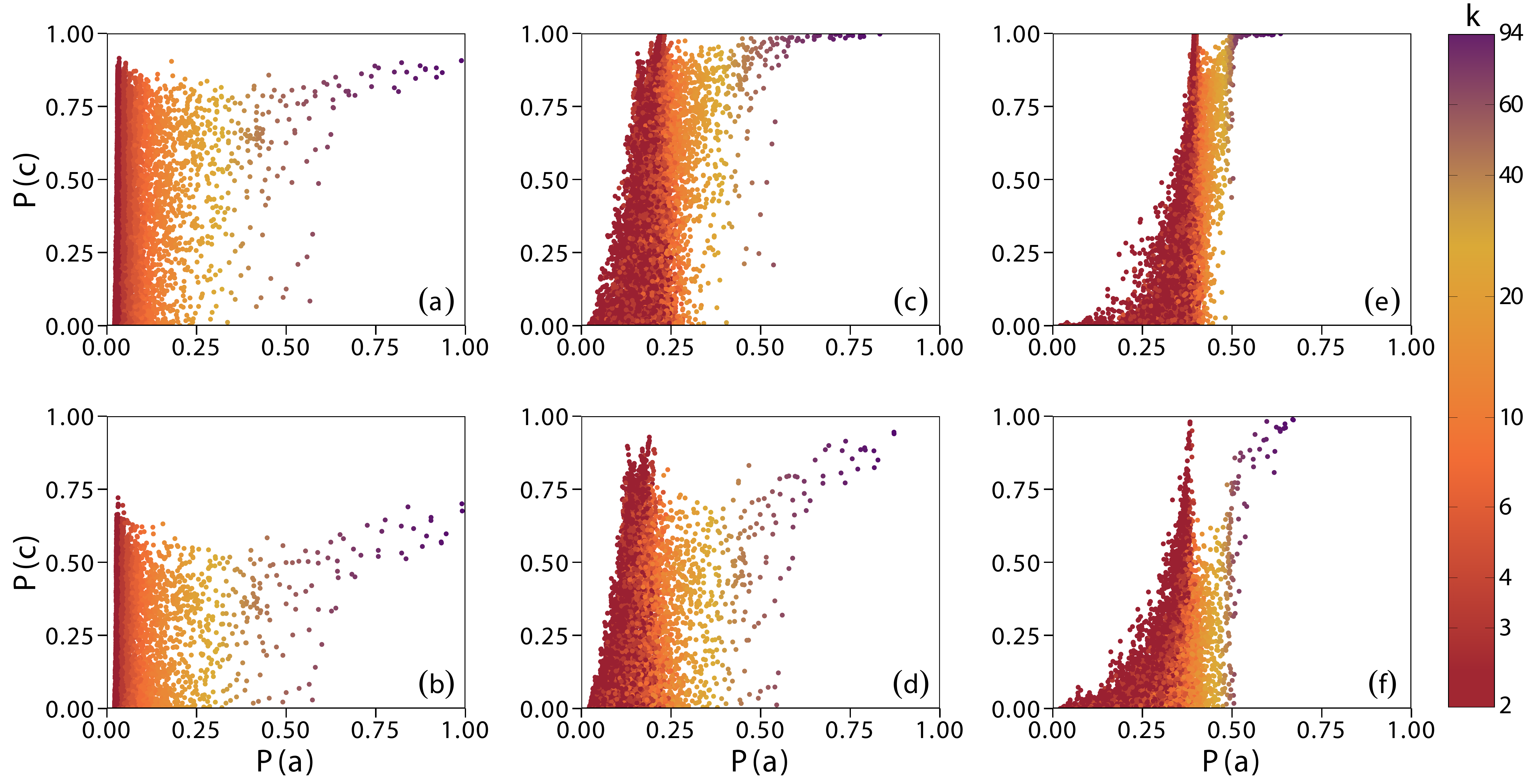,width=14cm}}
\caption{Cooperation probability in dependence on the activation probability for each player using payoff-correlated activity with different values of $\tau$ and two different values of the temptation to defect $b$ [(a) $\tau=1$, $b=2.4$; (b) $\tau=4$, $b=2.4$; (c) $\tau=8$, $b=2.4$; (d) $\tau=1$, $b=3.0$; (e) $\tau=4$, $b=3.0$; (f) $\tau=8$, $b=3.0$]. In addition, the degree of each player is color coded with the color bar on the right. For details regarding the interpretation of these results were refer to the main text.}
\label{both}
\end{figure*}

As results presented in Fig.~\ref{activate} show, this tradeoff is directly reflected in the activation probabilities of players when plotted against their degrees in the network. While the degree-correlated activity in the form $P_{a_i} = k_i/k_{max}$ where $k_{max}$ of course linearly interpolates between the player with the lowest and the highest degree, payoff-correlated activity increasingly deviates from this linear profile as $\tau$ increases in an S-shaped manner. Thus, both hubs and low-degree players become less active on the expense of average-degree hubs, which in turn see their activation probability increased. This transitions is the more pronounced the higher the value of the temptation to defect $b$, which also explains why the effects of the optimal value of $\tau$ are more pronounced at larger values of $b$. Nevertheless, the fact that hubs would be able to better maintain their cooperative state despite being active less of the time is contradictory, and opposes the argument we put forward for degree-correlated activity -- namely that it induces cascading failures of cooperation that are particularly severe on scale-free networks due to frequently inactive hubs. But looking at hubs alone as the backbone of cooperation is misleading. As recently emphasized by Morone and Makse \cite{morone_n15}, and shown explicitly for the evolution of cooperation in \cite{szolnoki_epl16}, not just the degree matters for the players to have a strong influence on the population, but also their neighborhoods. So even average-degree and low-degree players can be optimal influencers if they are surrounded by a hierarchy of high-degree hubs. As can be inferred from the networks depicted in Fig.\ref{nets}, and as is in general well-known, hubs in a scale-free network tend to be hierarchically linked, so that the hubs, which are also the oldest nodes, are preferentially linked with nodes of similar degree (and age). Thus, by supporting a healthy activity of average-degree players can be advantageous for the successful evolution of cooperation, even if it means the highest-degree nodes having a slightly lower activity in return, because the higher probability of cooperation of average-degree players helps the hubs to maintain cooperation too. In this argument, there is of course also an optimal tradeoff between just how much activity of high-degree nodes should be sacrificed for the higher activity of average-degree nodes, which in our case turns out to be warranted by $\tau=4$ in Eq.~(\ref{payc}), and which as shown in the inset of Fig.~\ref{optimal}, ensures the highest level of cooperation at sufficiently large values of $b$.

As the final result to corroborate our arguments, we show in Fig.~\ref{both} how the cooperation probability depends on the activation probability for two different values of $b$, for both of which the optimal value of $\tau$ is clearly inferable in Fig.~\ref{optimal}. For clarity, the degree of players with the corresponding values is also color coded via the color bar on the right of the figure. First, it can be observed that for $b=3.0$ (bottom three panels), the overall $P(c)$ values are lower than for $b=2.4$, which is expected because the higher severity of the social dilemma. Comparing the panels from left to right in both rows, however, shows quite clearly that for $\tau=4$ (middle), the activity probability of average-degree hubs is higher than for $\tau=1$, and these players also have a higher probability of cooperation. At the same time, hubs loose some of their activation, but their probability to cooperate goes up as well, which as we have argued, is driven by the revived status of the average-degree players. As $\tau=8$ the fast decay to baseline activity imposes a sharp barrier, which is particularly visible for $b=3.0$, which prevents the activity pattern to have a noticeable impact on the evolutionary outcome, effectively returning a result similar to the traditional version of the game where each player has a uniform activity.

\section{Discussion}

We have studied the impact of different activity patterns on the evolution of cooperation in the weak prisoner's dilemma game on scale-free networks. Through the introduction of different individual activation probabilities, we have considered degree-correlated and payoff-correlated activity patterns, with the later also having a tunable exponential decays to baseline activity. In both cases, we have assumed that active players play the game with their neighbors and obtain their payoffs accordingly, while inactive players do not play and thus end up with a null payoff. In agreement with this, only active players were able to pass their strategies to their neighbors, while inactive players were unable to do likewise. Within this setup, we have shown that degree-correlated activity introduces cascading failures of cooperation that are due to frequently inactive hubs. This in turn invokes the need for the spatiotemporal reorganization of compact cooperative clusters, which takes comparatively long, thus resulting in an overall lower cooperation level. Conversely, for payoff-correlated activity with an intermediately fast decay to baseline activity, we have demonstrated the existence of optimal conditions for the resolution of social dilemmas. In particular, we have shown that in this case the inactivity of hubs is compensated effectively by the increased activity of average-degree players. The sudden drops in the level of cooperation therefore no longer occur, and an overall higher level of cooperation in the population is attained.

We have further corroborated our main conclusions with the study of the average activity of players in dependence on their degree, showing that under the optimal conditions both the hubs and the low-degree players have a comparatively lower activity than under degree-correlated activity rule, while at the same time average-degree nodes have an elevated activity. There is transition from a linear activity profiles to and S-shaped activity profile in dependence on the degree. In this sense, the optimal result is on first glance contradictory, since, as is well known, cooperative hubs act as important centers of sizable and compact communities of cooperators on scale-free networks \cite{santos_prl05}. It would thus seem prohibitive that an even lower level of activity would improve the chances of success for cooperation. As it turns out, since the average-degree nodes are more active, and since they are predominantly linked directly to the hubs, their increased activity acts as a protector against a strategy change towards defection. Here the concept of collective influence, as recently introduced by Morone and Makse \cite{morone_n15}, becomes crucial, in that a more wholesome take on the problem that goes beyond an individual's degree and hub status is needed. In a nutshell, many average-degree, and even low-degree, nodes can be optimal influencers if they are ``surrounded by hierarchical coronas of hubs''. Thus, by supporting a healthy activity of average-degree players, even if it is on the expense of the hubs, ultimately turns out to be advantageous for sustenance of cooperation. Notably, a similar argument was recently introduced in \cite{szolnoki_epl16} to explain the optimal distribution of the potency of players for passing their strategies in a network.

We conclude by noting that even the relatively simple considerations of different activity patterns in evolutionary games reveal remarkable complexity in the underlying dynamics, involving cascading failures that present itself as asymmetric dynamic instabilities in the time course of the evolution of the two competing strategies under degree-correlated activity, and the progressive annihilation thereof under appropriately fine-tuned payoff-correlated activity. Although properly taking into account the diversity of individual activities of players \cite{capraro_pone13, capraro_sr14} is certainly an important aspect of bringing models closer to real-life conditions, it nevertheless seems a daunting proposition to add to the complexity of the studied model. In the future, it would be interesting to see experiments on human cooperation with an exit strategy, as studied before experimentally in the ultimatum game \cite{dana_07}, while modeling and simulation efforts could be directed towards relative times scales in evolutionary dynamics \cite{roca_prl06}, in that the typical time for an activation change is likely to be different from the time needed to make a strategy change. We hope that this paper will motivate further research along this line in the near future, as well as to draw attention to the importance of collective influence in networks for the evolution of cooperation in particular, and for social phenomena, ranging from efficient immunization to the diffusion of information in general.

\begin{acknowledgments}
A. A was supported by an FPI fellowship of MINECO, Spain. S. M. acknowledges support from Juan de la Cierva Program, Spain. M. P was supported by the Slovenian Research Agency (Grants J1-7009 and P5-0027). Y. M. acknowledges support from the Government of Arag\'on, Spain through a grant to the group FENOL, by MINECO and FEDER funds (grant FIS2014-55867-P) and by the European Commission FET-Proactive Project Multiplex (grant 317532).
\end{acknowledgments}

\end{document}